\documentclass{PoS}

\usepackage{cite,psfrag}

\title{Status of QCD}

\ShortTitle{Status of QCD}

\author{\speaker{Thomas Gehrmann}\\
        Institut f\"ur Theoretische Physik, Universit\"at Z\"urich, CH-8057 
Z\"urich\\
        E-mail: \email{thomas.gehrmann@uzh.ch}}

\abstract{In this talk, I
 review recent developments in perturbative QCD and their 
applications to collider physics.}

\FullConference{XVIII International Workshop on Deep-Inelastic Scattering and Related Subjects, DIS 2010\\
		April 19-23, 2010\\
		Firenze, Italy}

\begin{document}

\section{Introduction}
QCD is well established as theory of the strong interaction, and its perturbation theory expansion 
can be used to obtain quantitative predictions for observables in  high energy particle 
collisions. QCD effects are omnipresent in hadronic collisions, and a detailed 
understanding of QCD is mandatory for the interpretation of collider data, for new physics 
searches and for precision studies. In this talk, I review the recent progress on 
applications of QCD at high energy colliders.

\section{Jets and Event Shapes}
Hadronic jets are the final state signatures of quark or gluon production in particle collisions
at high energies. As such, they are important both as tool for precision studies of QCD, and 
in searches for new physics effects~\cite{Salam:2009jx}. 
Jets are defined through a jet algorithm (a measurement 
and recombination prescription to reconstruct the jet momenta from measured 
individual hadron momenta). Jet algorithms must fulfill infrared-safety criteria, i.e.\ the 
reconstructed jet kinematics must be insensitive to radiation of soft or collinear particles. 
Historically, two classes of jet algorithms were widely used at high energy colliders: cone-based algorithms
and sequential recombination algorithms. Cone-based algorithms allow an intuitive 
understanding of the jets, and can be formulated in an 
infrared-safe manner~\cite{Salam:2007xv}. 
Recombination algorithms are less intuitive, and their slow performance 
for a large number of final state particles was overcome only recently with the 
FastJet implementation~\cite{Cacciari:2005hq}. 
Variants of these algorithms differ in the distance measure 
used to identify neighboring momenta, it turns out the the so-called anti-$k_T$ 
recombination algorithm results in perfectly cone-shaped jets~\cite{Cacciari:2008gp}. 

Much recent progress has been made recently in using jets as analysis tools. The 
concept of the jet catchment area~\cite{Cacciari:2008gn} allows 
to obtain a geometrical interpretation of 
recombination algorithms, and to identify outside-jet regions, which can be used for 
underlying event studies. Aiming for the reconstruction of highly boosted massive particles,
the study of jet substructure~\cite{Butterworth:2008iy} 
has proven to be very promising. All decay products 
are first clustered in one fat jet, whose substructure is then resolved by lowering the 
resolution, resulting in a pronounced discontinuity once the particle decay is 
resolved. As one of the first results obtained
using this procedure, the reconstruction of $t\bar t H$ 
(a reaction that could not be observed with standard cut-based methods due to the 
large standard model background) final states appears to become feasible~\cite{Plehn:2009rk}.
 Many more applications are under study.  

Closely related to jet observables are event shapes, which characterize the geometrical 
properties of a hadronic final state. 
Distributions in several event shape variables  were measured very extensively by LEP
in view of precision studies of QCD. These results have a wide variety of applications, 
ranging from precision measurements of $\alpha_s$, tests of resummation, 
study of hadronization effects, and tuning of multi-purpose Monte Carlo event generators. 
At hadron colliders, event shapes were only studied little up to now, and their 
definition is more involved due to the restricted final state region usually accessible 
in this environment. If defined properly, they can serve as tools for model-independent 
searches~\cite{Banfi:2010xy}, and may be complementary to jet observables~\cite{Stewart:2010tn}.
An extensive classification of event shapes at hadron colliders has 
been made very recently~\cite{Banfi:2010xy}.

\section{Multiparticle Production at NLO}

The search for new physics signals at the CERN LHC will 
very often involve multi-particle final states, consisting of numerous jets, leptons, photons and 
missing energy. Quite in general, massive short-lived particles are detected 
through their decay signatures, as for example top quark pair production, which was 
first observed in final states with four jets, a lepton and missing energy.  

Meaningful searches for these signatures require not only a very good 
anticipation of the expected signal, but also of all standard model backgrounds
yielding identical final state signatures. Since leading-order calculations 
are affected by large uncertainties in their normalization and their 
kinematical dependence, it appears almost mandatory to include NLO corrections, 
which also allow to quantify the jet algorithm dependence, and effects of 
extra radiation. 
For a long time, these corrections were available only for at most 
three final state particles. 

An  NLO calculation of a $n$-particle observable consists of two contributions: the virtual 
one-loop correction to the $n$-particle production process, and the real radiation contribution 
from the $(n+1)$-particle production process. Both contributions are infrared divergent, and 
can be evaluated numerically only after extracting the infrared divergent contributions 
from the real radiation process. Several  well-established and widely used methods exist for 
this task~\cite{Frixione:1995ms,Catani:1996vz,Catani:2002hc,Kosower:1997zr,Campbell:1998nn,Daleo:2006xa,GehrmannDeRidder:2009fz,Somogyi:2009ri}. 

The evaluation 
of the one-loop multi-leg amplitudes poses a challenge in complexity (due to the large 
number of diagrams, and large number of different scales present) and stability (due to 
possible linear dependences among the external momenta). It has been known for long that 
any one-loop amplitude can be expressed as a linear combination of one-loop integrals 
with at most four external legs, plus a rational remainder. Enormous progress has been made
in recent years in the systematic computation of the one-loop integral coefficients and 
rational terms. While previously established
Feynman-diagram based techniques for tensor reduction and form factor 
decomposition were successfully extended~\cite{Denner:2005nn,Binoth:2005ff} to 
multi-leg problems, a new arsenal of techniques was emerging from the use of 
unitarity and multi-particle cuts~\cite{Bern:1994zx}. Using these, the 
one-loop integral coefficients of an amplitude can be inferred~\cite{Britto:2004nc,Britto:2006sj,Mastrolia:2006ki,Forde:2007mi} without evaluation of all 
individual diagrams. An extension of these ideas is made by performing the reduction 
at the integrand level in the OPP method~\cite{Ossola:2006us,Ossola:2008xq}. The rational
 coefficients can be determined in the same framework by extending the unitarity relations 
 from four dimensions to higher-dimensional space-time~\cite{Ellis:2007br,Giele:2008ve,Ellis:2008ir}.
 
Given the large number of different multi-particle final states of potential interest to 
new physics searches, an automation of NLO calculations is highly desirable. Based 
on existing multi-purpose leading order matrix element generators, the implementation of 
the real radiation contributions and their infrared subtraction terms is straightforward, and has 
been accomplished in the Sherpa~\cite{Gleisberg:2007md}, 
MadGraph~\cite{Frederix:2008hu,Frederix:2010cj,Frederix:2009yq} 
and Helac/Phegas~\cite{Czakon:2009ss}
 frameworks, as well as in the form of 
independent libraries~\cite{Seymour:2008mu,Hasegawa:2009tx}, which 
complement already existing libraries
 in the MCFM~\cite{Campbell:1999ah,Campbell:2002tg} and 
 NLOJET++~\cite{Nagy:2003tz}
packages. The automation of the virtual corrections is a much larger challenge, 
which is currently being accomplished in several program packages based on 
the various available methods. A semi-numerical form factor decomposition is automated 
in the Golem package~\cite{Binoth:2008uq}. Unitarity and multi-particle cuts are used in 
the BlackHat package~\cite{Berger:2008sj}, and the OPP method is automated in 
CutTools~\cite{Ossola:2007ax}. Numerical $D$-dimensional unitarity is applied in 
the Rocket package~\cite{Giele:2008bc}
and the Samurai package~\cite{Mastrolia:2010nb}; it also forms the 
basis of several currently ongoing implementations~\cite{Giele:2009ui,Lazopoulos:2009zn}.

Several NLO calculations of $2\to 3$ reactions at hadron colliders were completed recently. 
These include the production of two vector bosons and one 
jet~\cite{Dittmaier:2007th,Dittmaier:2009un,Campbell:2007ev,Binoth:2009wk,Campanario:2009um,Campanario:2010hp}, of a Higgs boson and two jets~\cite{Campbell:2006xx,Badger:2009hw,Badger:2009vh,Campbell:2010cz},
of $t\bar t Z$~\cite{Lazopoulos:2008de}, 
and of three vector bosons~\cite{Lazopoulos:2007ix,Binoth:2008kt,Hankele:2007sb,Campanario:2008yg,Bozzi:2009ig}. Of a similar kinematical type are vector boson fusion processes,
which are computed to NLO accuracy in the VBFNLO package~\cite{Arnold:2008rz}. 
The current frontier of complexity are NLO calculations of $2\to 4$ reactions. Several 
very important processes of this type have been computed most recently. 

An important channel for Higgs boson searches, and for 
subsequent determinations of Yukawa couplings, is the associated production of a Higgs with a 
heavy quark-antiquark pair, with the Higgs boson decaying into $b\bar b$. The QCD background 
processes yielding $t\bar t b\bar b$ final states were computed recently to 
NLO~\cite{Bredenstein:2008zb,Bredenstein:2009aj,Bredenstein:2010rs,Bevilacqua:2009zn}, 
displaying
 moderate but non-constant QCD corrections, which show a  non-trivial dependence on the 
 event selection cuts. A natural extension of these calculations are $t\bar t+2j$ final 
 states~\cite{Bevilacqua:2010ve}. 
 Extended Higgs sectors predict a sizable rate of associated production with bottom quark 
 pairs, and the calculation of $b\bar b b\bar b$ final states is in progress~\cite{Binoth:2009rv}. 
\begin{figure}[t]
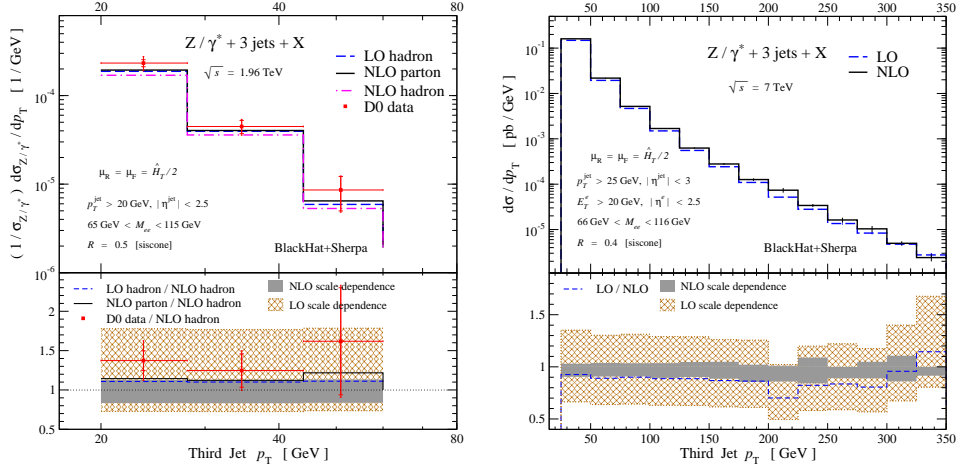

\centerline{
\includegraphics[width=0.4\textwidth]{Z3j-D0-HT_siscone-Pt20_jets_jet_1_3_pt_3_with_D0_data.eps}\hspace{4mm}
\includegraphics[width=0.4\textwidth]{Z3j-7TeV-HTp_siscone-R4-Pt25_jets_jet_1_1_pt__D3.eps}}
\caption{Transverse momentum distribution of the third-hardest jet in $Z^0+3j$ events 
at the Tevatron (left) and the LHC (right) to LO and NLO.
Figure taken from~\protect{\cite{Berger:2010vm}}. }\label{fig:z3j}
\end{figure}

The final state signature of a vector boson and three hadronic jets is often relevant in 
generic new physics searches. NLO corrections of $W+3j$ were obtained by two groups in the 
Rocket~\cite{KeithEllis:2009bu} and in the 
Blackhat+Sherpa~ \cite{Berger:2009zg,Berger:2009ep} framework. 
The corrections to $Z^0+3j$ were also 
obtained with Blackhat+Sherpa~\cite{Berger:2010vm}. 
For both observables,  corrections are moderate, and stabilize the 
QCD prediction to the ten per cent level, required for precision phenomenology, as 
can be seen in Figure~\ref{fig:z3j} from~\cite{Berger:2010vm}. 
Knowledge of the NLO corrections to these processes 
allows many phenomenological studies, such as for example the stability of  final state 
correlations~\cite{Berger:2009ep} 
under perturbative corrections, and the optimal choice of scales in multi-scale 
processes~\cite{Berger:2009ep,Berger:2010vm,Melnikov:2009wh,Bauer:2009km}. A crossing of 
$Z^0+3j$ is the process $e^+e^-\to 5j$, which was measured at LEP. The NLO 
calculation of it is in progress.

\section{Precision Observables at NNLO}
Few benchmark observables 
(e.g.\ jet cross sections, vector boson production, heavy quark production) 
are measured experimentally to an accuracy of one per cent or 
below. For a theoretical interpretation of these observables, an NLO description  (which 
has a typical residual uncertainty around ten per cent) is insufficient: extractions of 
fundamental parameters from these observables would be limited by the 
theory uncertainty. For a meaningful interpretation of these observables, NNLO corrections
are mandatory. Likewise, NNLO corrections are required for 
a reliable description of observables with 
potentially large perturbative corrections, like Higgs boson production.

The calculation of NNLO corrections to an $n$-particle final state
requires three ingredients: the two-loop matrix elements for the $n$-particle production, 
the one-loop matrix elements for the $(n+1)$-particle production and the tree-level
matrix elements for $(n+2)$-particle production. The latter two contributions develop 
infrared singularities if one or two particles become soft or collinear, requiring a 
subtraction method to extract these infrared poles, which are then combined with the 
virtual corrections to yield a finite prediction. The two major challenges of NNLO calculations 
are the two-loop matrix elements and the handling of the real 
radiation at NNLO. Up to now, two types of approaches to real radiation have been applied 
in NNLO calculations of exclusive observables. The sector decomposition 
method~\cite{Binoth:2000ps,Anastasiou:2003gr,Binoth:2004jv}
is based on a systematic expansion in distributions, followed by numerical integration over 
many different small phase space sectors. Subtraction methods search to approximate 
the full real radiation contribution by subtraction terms in all unresolved limits; these terms 
are then integrated analytically. While many subtraction methods have been worked out at NLO,
only two methods have so far yielded results at NNLO: the antenna subtraction 
method~\cite{GehrmannDeRidder:2005cm}
for processes in $e^+e^-$ annihilation, and the $q_T$-subtraction~\cite{Catani:2007vq} for 
hadron collider processes in specific kinematic configurations. Alternative approaches 
are under intensive development~\cite{Somogyi:2008fc,Aglietti:2008fe,Bolzoni:2009ye}. 
A combination of subtraction with sector decomposition~\cite{Czakon:2010td}
may hold the potential to become a general multi-purpose method.

The dominant Higgs boson production process is gluon fusion, mediated through a top quark 
loop. 
This process has been computed (in the infinite top mass limit) to NNLO
accuracy in a fully exclusive form including 
the Higgs boson decay, i.e.\ allowing for arbitrary infrared-safe final state cuts, 
both using sector decomposition~\cite{Anastasiou:2004xq,Anastasiou:2005qj,Anastasiou:2007mz}  
and using $q_T$-subtraction~\cite{Grazzini:2008tf,deFlorian:2009hc}.  These results 
can be directly applied to the Higgs boson search at the Tevatron, based on a neural
network combination of many different kinematical distributions~\cite{Anastasiou:2009bt}.
 Finite top mass effects 
at NNLO were derived most recently~\cite{Harlander:2009mq,Harlander:2009my,Pak:2009dg}
for the inclusive gluon fusion cross section. 
At this level of precision, 
mixed QCD and electroweak corrections~\cite{Anastasiou:2008tj} 
become equally important. The gluon 
fusion reaction can be mediated through loops involving any type of 
massive  color-charged particles, thereby offering an indirect constraint on physics
beyond the standard model, such as 
supersymmetric particles~\cite{Harlander:2003bb,Harlander:2004tp,Degrassi:2008zj,Anastasiou:2008rm,Muhlleitner:2008yw}, extra heavy quark 
families~\cite{Anastasiou:2010bt} or color-octet scalars~\cite{Boughezal:2010ry}.

Another very promising Higgs discovery channel is vector boson fusion. 
The factorizable NNLO corrections to the inclusive cross section for this process are 
closely related to inclusive deep inelastic scattering. They were computed very 
recently~\cite{Bolzoni:2010xr}, and 
turn out to be rather small, resulting in a high theoretical stability of the prediction. This 
channel can be equally sensitive on supersymmetric contributions~\cite{Hollik:2008xn}. 
\begin{figure}[t]
\centerline{
\includegraphics[width=0.45\textwidth]{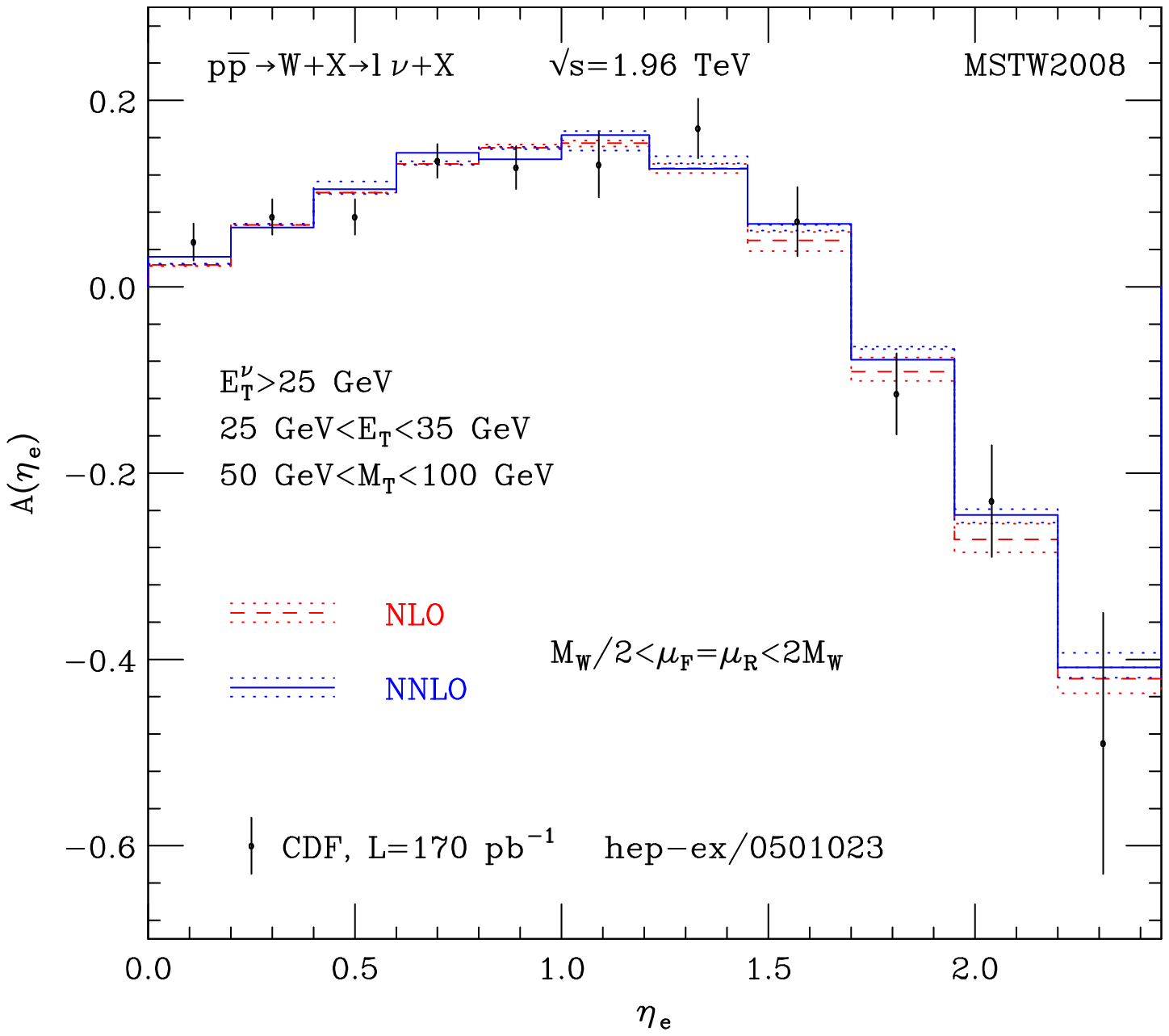}
\includegraphics[width=0.45\textwidth]{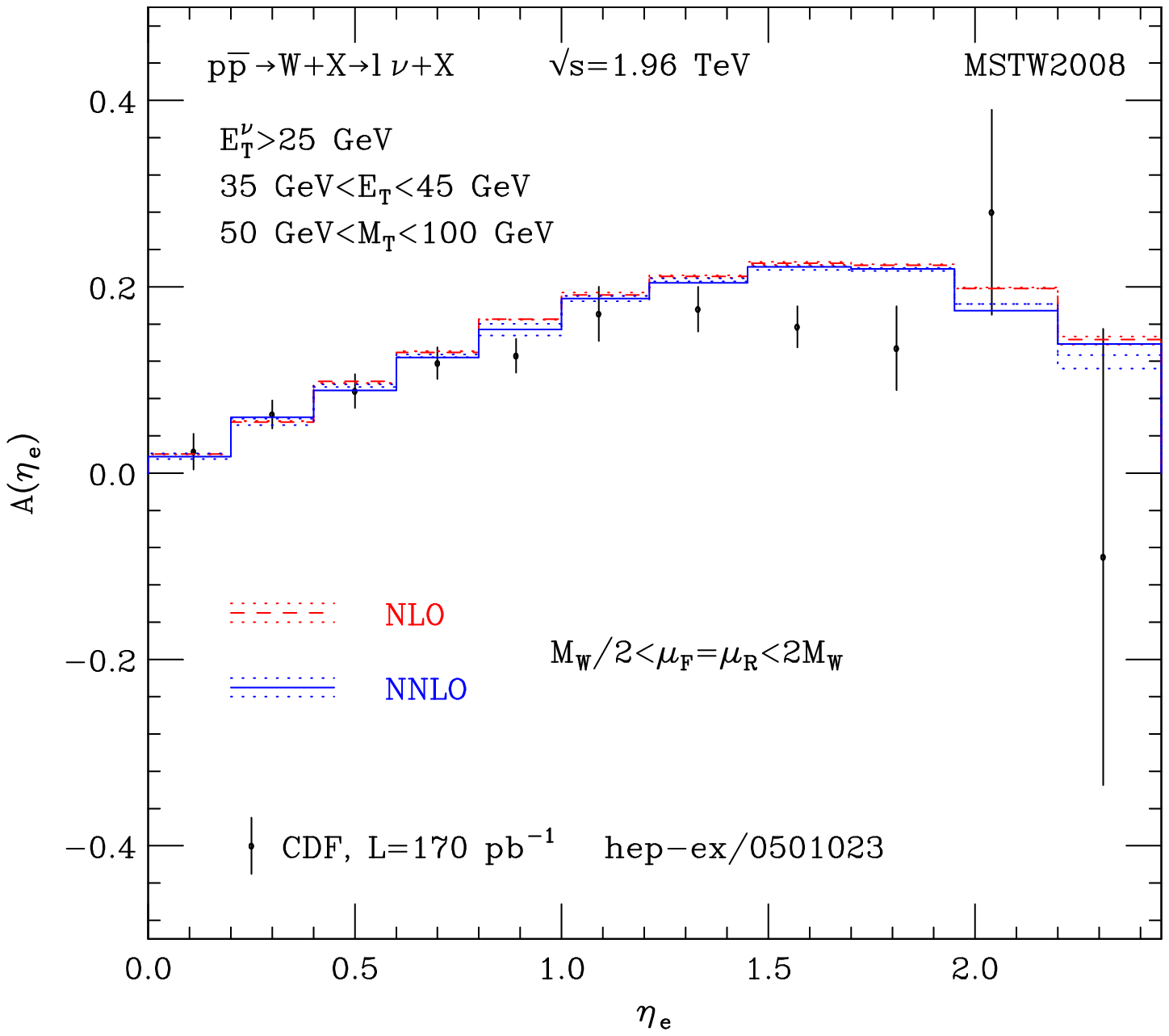}}
\caption{Lepton charge asymmetry at the Tevatron at NLO and NNLO, compared
to CDF data. Figure taken from~\protect{\cite{Catani:2010en}}. }\label{fig:wch}
\end{figure}

Fully exclusive NNLO corrections to vector boson production have equally been derived 
using sector decomposition~\cite{Melnikov:2006di,Melnikov:2006kv} 
and with $q_T$-subtraction~\cite{Catani:2009sm}, including the leptonic vector boson
decay. Observables derived from vector boson production are very important for precision 
studies of the electroweak interaction, and for the determination of the quark distributions in 
the proton. Using the newly obtained results, the NNLO corrections 
(and their uncertainty) to the lepton charge 
asymmetry~\cite{Catani:2010en} 
can be quantified, see Figure~\ref{fig:wch}, and this observable can be 
consistently included into NNLO fits of parton distributions.

\begin{figure}[t]
\centerline{\includegraphics[angle=-90,width=0.80\textwidth]{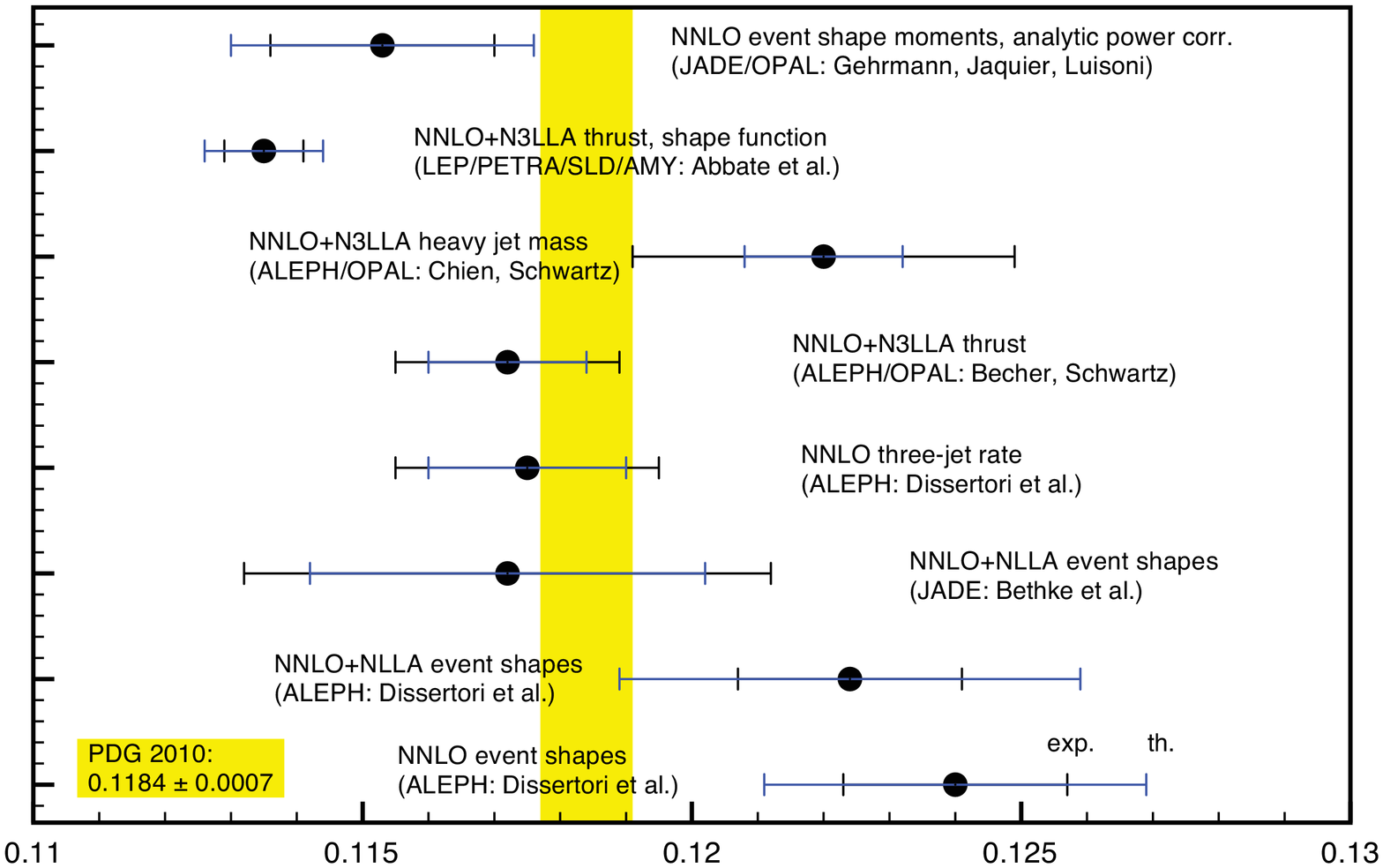}}
\caption{Determinations of $\alpha_s$ from event shapes and jet cross sections 
in $e^+e^-$ annihilation at NNLO, compared to the Particle Data Group 
world average. Experimental errors are indicated in black, theoretical errors in blue.}\label{fig:as}
\end{figure}

Jet production observables have been computed to NNLO only for $e^+e^-$ annihilation 
up to now. Two implementations of the NNLO corrections to $e^+e^-\to 3j$ and 
related observables are available~\cite{GehrmannDeRidder:2007jk,GehrmannDeRidder:2007hr,GehrmannDeRidder:2008ug,GehrmannDeRidder:2009dp,Weinzierl:2008iv,Weinzierl:2009ms,Weinzierl:2009nz,Weinzierl:2009yz},
both based on antenna subtraction. The 
magnitude of the NNLO corrections 
differs substantially between different event shape observables; including these new NNLO 
corrections, LEP data on event shapes and jet cross sections were reanalyzed
in view of an improved determination of the strong coupling constant. In general, an 
improved consistency among different observables was observed. To use 
measurements over an extended kinematical range, resummation of large logarithmic corrections 
in the two-jet limit is needed. This is available to next-to-leading logarithmic accuracy (NLLA) 
for all shape variables~\cite{Catani:1992ua,Gehrmann:2008kh}, 
and to N$^3$LLA for thrust~\cite{Becher:2008cf,Abbate:2010xh}
 and heavy jet mass~\cite{Chien:2010kc} distributions. The by-now
limiting factor in precision physics with event shape observables in $e^+e^-$ annihilation 
is the description of the parton-to-hadron transition (hadronization), which was previously 
modeled from parton shower based event generators. Substantial differences are observed 
between the LEP-era programs and more modern generators, and to analytic 
approaches to hadronization, 
based on the shape function formalism~\cite{Becher:2008cf,Abbate:2010xh,Chien:2010kc} 
and on a dispersive model~\cite{Dokshitzer:1995qm,Davison:2008vx,Gehrmann:2009eh}.  
The recent determinations of the strong coupling constant from event shapes and jet cross 
sections at NNLO~\cite{Becher:2008cf,Abbate:2010xh,Chien:2010kc,Gehrmann:2009eh,Bethke:2008hf,Dissertori:2009ik,Dissertori:2009qa,Dissertori:2007xa}
are summarized in Figure~\ref{fig:as}. Electroweak NLO corrections 
to jet observables~\cite{Denner:2009gx,Denner:2010ia,Denner:2009gj} are potentially  
of the same numerical importance as NNLO QCD corrections, and could be included in 
future studies.

In view of the very precise jet production data from 
HERA~\cite{turcato,krueger} and the Tevatron~\cite{li,casarsa}, the derivation 
of NNLO corrections to jet cross sections in  hadronic collisions is of high priority. The 
relevant two-loop matrix elements for hadronic collisions and for 
deep inelastic scattering~\cite{Gehrmann:2009vu}
are known for some time already, and substantial progress 
is being made to extend the antenna subtraction method to include hadrons in the initial state. 
The proper functioning of this method on the $gg\to 4g$ subprocess to hadronic dijet 
production has been demonstrated~\cite{Glover:2010im} most recently. 
The integrated forms of all antenna 
functions have been derived for one parton in the initial state~\cite{Daleo:2009yj}, the case 
of two initial state partons~\cite{Boughezal:2010ty} is work in progress. 

The large number of top quark pairs expected to be produced at the LHC will allow for 
precision top quark studies, requiring NNLO accuracy on the theoretical side. 
The relevant two-loop matrix elements were first derived in the high energy 
limit~\cite{Czakon:2007ej,Czakon:2007wk}. The 
exact $q\bar q\to t\bar t$ matrix element is known numerically~\cite{Czakon:2008zk}, 
substantial parts of it have been 
confirmed by an analytic calculation~\cite{Bonciani:2008az,Bonciani:2009nb}. 
The one-loop self-interference 
contributions are  equally known~\cite{Korner:2008bn,Anastasiou:2008vd,Kniehl:2008fd} 
The matrix elements with one and two extra partons
form part of the $t\bar t+j$ production at 
NLO~\cite{Dittmaier:2007wz,Dittmaier:2008uj,Melnikov:2010iu}. Methods 
to handle real radiation at NNLO in the presence of massive top quarks are currently under 
intensive development. Generalizing the subtraction method 
of~\cite{Frixione:1995ms} to NNLO and 
numerically integrating the relevant subtraction terms using sector 
decomposition~\cite{Czakon:2010td}
may provide a powerful method by combining the virtues of both approaches.

\section{Parton Distributions}
The parton distribution functions in the proton are a crucial ingredient in all 
hadron collider cross sections. They are determined (at LO, NLO or NNLO) from global 
fits~\cite{forte}
to a variety of 
data sets from fixed target experiments, from HERA and from the Tevatron.
On the theory side, the parton distribution fits require the DGLAP splitting 
functions (which govern the scale evolution of the parton distributions, and are known to 
NNLO~\cite{Moch:2004pa,Vogt:2004mw})
and coefficient functions for each process considered in the global fit. These 
coefficient functions are known to NNLO for inclusive DIS~\cite{Zijlstra:1992qd}, 
the Drell-Yan process~\cite{Melnikov:2006kv} and for 
heavy quark production in DIS~\cite{Bierenbaum:2009mv}, 
but only to NLO for jet production observables. The 
fit procedure must incorporate experimental and theoretical errors in a consistent manner. 

Global fits of parton distributions are performed by various collaborations, with slight differences 
in the methodology~\cite{forte}. Recent sets of parton distributions  are from 
MSTW~\cite{Martin:2009iq}, CTEQ~\cite{Pumplin:2009nk}, 
JR~\cite{JimenezDelgado:2008hf}, NNPDF~\cite{Ball:2010de} and 
ABKM~\cite{Alekhin:2009ni}. A comparison of them shows that the quark distributions are 
known rather precisely at large $x$, while the gluon distribution is uncertain to within ten per cent 
at large $x$, and systematic differences between the fits exist within errors. For small values of
$x<10^{-3}$, uncertainties on the distributions become very large.

\section{Infrared Structure and Resummation}
The perturbative expansion of QCD observables in the strong coupling constant is 
reliable if only a single hard scale is present, it becomes problematic for observables 
depending on several hard scales, leading to large logarithmic corrections at all orders. 
In these cases, a rearrangement of the perturbative series by means of a resummation of 
large logarithmic corrections often appears more suitable. 

Resummation of leading logarithmic corrections is accomplished by event generators~\cite{richardson} based 
on parton showers, initially based on leading order calculations.
Parton showers can be combined with fixed order NLO calculations in the 
MC@NLO~\cite{Frixione:2002ik} or the POWHEG~\cite{Frixione:2007vw} approach. 
The MC@NLO event generator already covers 
a large number of different processes, with $W^\pm t$ production~\cite{White:2009yt} 
and $H^\pm t$ production~\cite{Weydert:2009vr}
among the most recent additions. Within POWHEG, single top production~\cite{Alioli:2009je} 
and Higgs production 
in vector boson fusion~\cite{Alioli:2008tz} were accomplished 
most recently. The POWHEG box~\cite{Alioli:2010xd} provides 
users with a 
framework for implementing existing NLO calculations in this framework. 

A detailed understanding of the infrared structure of QCD can be gained from the 
observation that infrared poles in loop amplitudes translate into large logarithms in real 
radiation processes and vice versa. This relation can be applied successfully in both directions: 
for example to
predict infrared poles at two loops from resummation~\cite{Catani:1998bh,Sterman:2002qn} 
and to extract large-$x$ resummation 
constants~\cite{Moch:2005ba} from the poles of the QCD form factors. 
By relating the infrared poles in QCD to 
ultraviolet poles in soft-collinear effective theory (SCET)~\cite{Bauer:2000yr}, 
it becomes possible to express the infrared pole structure of 
QCD amplitudes by a multiplicative renormalization in SCET. 
Based on constraints~\cite{Becher:2009cu,Gardi:2009qi} and symmetry 
arguments, it becomes 
possible to  conjecture that the infrared pole structure of massless  QCD 
multi-loop amplitudes 
is uniquely determined~\cite{Gardi:2009qi,Becher:2009qa,Becher:2009cu,Dixon:2009ur}
by the cusp anomalous dimension and the collinear anomalous 
dimensions of the external particles.

The resummed description of an observable consists~\cite{Sterman:1986aj,Catani:1989ne}
of a hard coefficient, a soft function, 
jet functions containing final state collinear radiation and parton distributions containing 
initial state collinear radiation. In SCET~\cite{Bauer:2000yr}, 
each of these elements is identified with an operator or
a non-local function. The resummation~\cite{Idilbi:2006dg,Becher:2006mr}
then proceeds by computing their anomalous dimensions
and solving the respective evolution equations. First applications of SCET-based resummation 
are the thrust~\cite{Becher:2008cf,Abbate:2010xh}
and heavy jet mass~\cite{Chien:2010kc} distributions in $e^+e^-$ annihilation, the 
inclusive Drell-Yan and Higgs production~\cite{Idilbi:2006dg,Ahrens:2008nc}
and inclusive photon production~\cite{Becher:2009th}. This topic is 
currently under fast development, any many yet open issues, like jet production and radiation 
off incoming partons~\cite{Stewart:2009yx,Stewart:2010qs}
are being addressed.

\begin{figure}[t]
\centerline{
\psfrag{x}[][][1][90]{$d\sigma/dM$ [fb/GeV]}
\psfrag{y}[]{$M$ [GeV]}
\psfrag{z}[][][0.85]{$\sqrt{s}=1.96$\,TeV}
\includegraphics[width=0.45\textwidth]{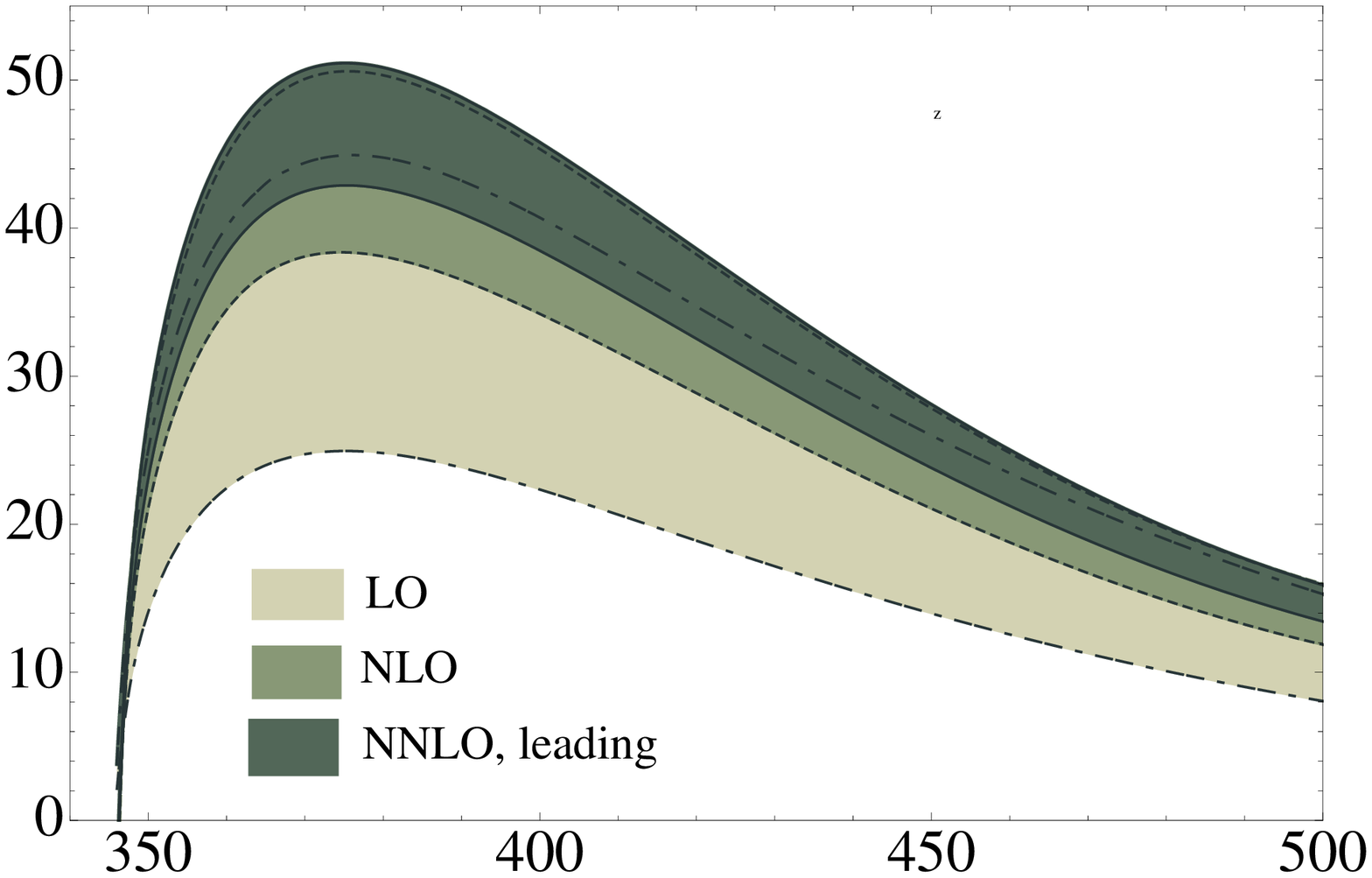}
\psfrag{x}[][][1][90]{$d\sigma/dM$ [fb/GeV]}
\psfrag{y}[]{$M$ [GeV]}
\psfrag{z}[][][0.85]{$\sqrt{s}=1.96$\,TeV}
\includegraphics[width=0.45\textwidth]{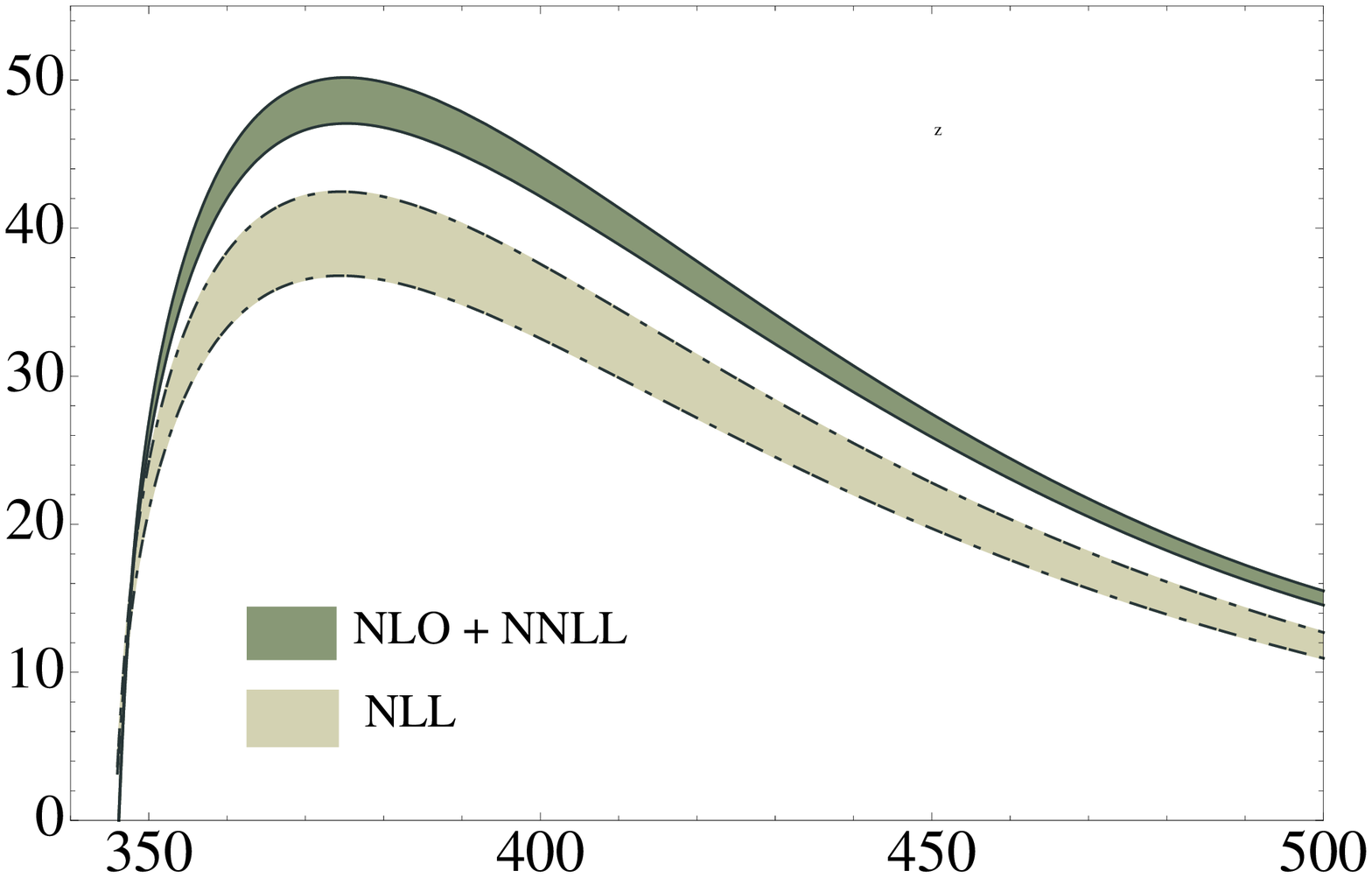}}
\caption{Top quark invariant mass distribution at the Tevatron in fixed order 
expansion (left) and resummation (right).
Figure taken from~\protect{\cite{Ahrens:2010zv}}. }\label{fig:ttbarresum}
\end{figure}

Many of the constraints used to obtain the all-order conjecture for massless QCD amplitudes 
do not apply in the presence of particle masses. Consequently, the pole structure of massive 
amplitudes is more involved; in particular, it contains multi-particle 
correlations~\cite{Ferroglia:2009ii}, which were 
absent in the massless case. Only recently, a prediction of the infrared poles to two-loop order 
has been accomplished~\cite{Ferroglia:2009ii,Beneke:2009rj}.
With these results, the resummation of the top quark pair production 
cross section to third logarithmic order (NNLL) could be completed. While dominant 
contributions  at this order were known for some 
time~\cite{Moch:2008qy,Kidonakis:2008mu} 
the full corrections have been obtained now
in two approaches:  
based massive soft anomalous dimensions~\cite{Czakon:2009zw,Czakon:2008cx} 
and by using SCET~\cite{Ahrens:2010zv}. The $t\bar t$ invariant mass 
distribution is compared in fixed order and  resummed expansion in 
  Figure~\ref{fig:ttbarresum}, taken from~\cite{Ahrens:2010zv}. It can be seen that 
  the resummation has only moderate numerical impact on the central value, but results in 
  a substantial reduction of the scale uncertainty.  By expanding 
the resummed  results to fixed order, one can in turn approximate the NNLO corrections to 
the top quark production cross section~\cite{Moch:2008qy,Beneke:2009ye,Aliev:2010zk}.

\section{Multi-loop Results}
Several reasons motivate the derivation of perturbative corrections even beyond NNLO. 
Such theoretical accuracy is 
required to describe very precisely measured quantities (like sum rules or total decay rates)
or for processes with a very slowly converging perturbative series. Moreover, these corrections 
allow insight into the infrared structure of QCD at high orders, and determine 
resummation coefficients.  

The current frontier of complexity for QCD loop amplitudes are $2\to 4$ processes at one loop, 
$2\to 2$ processes at two loops, $1\to 2$ processes at three loops and $1\to 1$ processes at 
four loops. Various innovative techniques have allowed substantial progress on 
multi-loop QCD calculations in recent times: a substantial reduction of complexity 
is achieved by reducing the large number of integrals appearing in a calculation 
to a small number of master integrals by exploiting linear relations among the 
integrals~\cite{Chetyrkin:1981qh}. 
Various techniques have proven successful in the derivation of these master integrals:
for example the Mellin-Barnes transformation~\cite{Smirnov:1999gc,Tausk:1999vh}, 
differential equations~\cite{Kotikov:1991hm,Kotikov:1991pm,Remiddi:1997ny,Gehrmann:1999as} 
in masses and momenta, 
and the sector decomposition technique~ \cite{Binoth:2000ps,Smirnov:2008py}. 
The reduction to master integrals is usually based on a lexicographic ordering
(the Laporta algorithm~\cite{Laporta:2001dd}). 
Implementations of this algorithm are available in several computer 
algebra frameworks: the AIR package~\cite{Anastasiou:2004vj} in Maple, 
the FIRE package~\cite{Smirnov:2008iw} in 
Mathematica, and the Reduze package~\cite{Studerus:2009ye}
 as a stand-alone C++ implementation based on 
Ginac and CLN. 

The Mellin-Barnes method allows to express master integrals in a systematic 
manner~\cite{Czakon:2005rk} in a form suitable for analytical or numerical evaluation. 
Recent results 
obtained with this method include the massless three-loop QCD form factor 
integrals~\cite{Heinrich:2007at,Heinrich:2009be,Lee:2010cg}. Another very 
successful method is the glue-and-cut technique, which 
exploits relations among topologically different master integrals. It recently led 
to the derivation of the four-loop propagator master integrals~\cite{Baikov:2010hf}. Many of these 
results were validated independently~\cite{Smirnov:2010hd} using the sector 
decomposition technique. 

The quark and gluon form factors are the simplest infrared-divergent objects in QCD. They 
allow the determination of resummation coefficients and enter benchmark reactions like 
Drell-Yan and Higgs production. They were recently computed to three 
loops~\cite{Baikov:2009bg,Gehrmann:2010ue}. 
The static QCD  potential formed by an infinitely heavy quark-antiquark pair is an important 
ingredient in the determination of heavy quark masses from sum rules, and in the description 
of top quark pair production at threshold. It was computed at three-loop accuracy most 
recently~\cite{Anzai:2009tm,Smirnov:2009fh}. 

Many important observables can be expressed as two-point functions: total decay rates, 
sum rules and moments of structure functions. Massless two-point functions were obtained 
recently at four loops. Among the results derived in this context are the hadronic 
$R$-ratio and the $\tau$-decay rate~\cite{Baikov:2008jh}, which allow some of the most accurate 
 determinations of the strong coupling constant: $\alpha_s(M_Z)_R = 0.1190\pm 0.0026$
 and $\alpha_s(M_Z)_\tau = 0.1202\pm 0.0019$. Most recent results~\cite{Baikov:2010je}
 are the polarized Bjorken sum rule, the Adler function and the Crewther relation.

\section{Conclusions}
 QCD is crucial for the success of the LHC physics programme in understanding 
 signals and backgrounds, knowing parton distribution functions, and using jets and 
 event shapes as analysis tools. Particle theory is getting ready for this challenge on 
 many frontiers: with improved jet algorithms and event shape definitions, with an 
 enormous progress on NLO calculations for multi-leg final states, with first NNLO 
 results for precision observables, with an emerging understanding of the all-order structure of
 infrared singularities, and with landmark results at three loops and beyond. 

\section*{Acknowledgments}
The author gratefully ackowledges the 
 support by the Swiss National Science Foundation
(SNF) under  contract 200020-126691.

\end{document}